 \newlength\smallfigwidth
 \documentclass[aps,prb,twocolumn,floatfix,showpacs,amsmath,amssymb]{revtex4}

\usepackage{graphicx}
\usepackage[hypertex]{hyperref}
\begin{document}

\preprint{UFV/UFJF}

\title{Predicted defect induced vortex core switching in thin magnetic nanodisks}
\author{R.\ L.\  Silva}
\email{ricardodasilva@ufv.br} \affiliation{ Departamento de
F\'isica, Universidade Federal de Vi\c cosa, Vi\c cosa, 36570-000,
Minas Gerais, Brazil }
\author{N.\ M.\ Oliveira-Neto}
\affiliation{Departamento de F\'isica, Universidade Estadual so
Sudoeste da Bahia, 45200-000, Jequi\'{e}, Bahia, Brazil}
\author{S.\ A.\ Leonel}
\affiliation{ Departamento de F\'isica ICE, Universidade Federal
de Juiz de Fora, Juiz de Fora 36036-330, Minas Gerais, Brazil }
\author{P.\ Z.\ Coura}
\affiliation{ Departamento de F\'isica ICE, Universidade Federal
de Juiz de Fora, Juiz de Fora 36036-330, Minas Gerais, Brazil }
\author{A.\ R.\ Pereira}
\email{apereira@ufv.br} \affiliation{ Departamento de F\'isica,
Universidade Federal de Vi\c cosa, Vi\c cosa, 36570-000, Minas
Gerais, Brazil }
\author{W.\ A.\ Moura-Melo}
\email{winder@ufv.br} \affiliation{ Departamento de F\'isica,
Universidade Federal de Vi\c cosa, Vi\c cosa, 36570-000, Minas
Gerais, Brazil }
\author{R.\ C.\ Silva}
\affiliation{ Departamento de F\'isica, Universidade Federal de
Vi\c cosa, Vi\c cosa, 36570-000, Minas Gerais, Brazil }

\date{Fev. 13, 2007}

\begin{abstract}
We investigate the influence of artificial defects (small holes)
inserted into magnetic nanodisks on the vortex core dynamics. One
and two holes (antidots) are considered. In general, the core
falls into the hole but, in particular, we would like to remark an
interesting phenomenon not yet observed, which is the vortex core
switching induced by the vortex-hole interactions. It occurs for
the case with only one hole and for very special conditions
involving the hole size and position as well as the disk size. Any
small deformation in the disk geometry such as the presence of a
second antidot changes completely the vortex dynamics and the
vortex core eventually falls into one of the defects. After
trapped, the vortex center still oscillates with a very high
frequency and small amplitude around the defect center.

\end{abstract}

\maketitle
\section{Introduction and motivation}

\indent As it is well known, magnetic vortex states are
experimentally observed in ferromagnetic disk-shaped
nanostructures. These topological objects exhibit a planar-like
configuration of spins outside the core, where a perpendicular
magnetization (up or down polarization) is observed
\cite{vortex,v2,v3}. As long as one could manipulate these
vortices, several possibilities would emerge. Therefore, studies
taking into account the application of external potentials such as
static and sinusoidal magnetic fields are nowadays very common in
the literature
\cite{Pulwey01,Volodymyr07,Xiao07,Yamada07,Novosad05,Guslienko02}.
For instance, the simplest effect induced by an external field is
the gyrotropic mode, which is the lowest excitation of the vortex
structure. This mode is simply the elliptical vortex core motion
(with the resonance frequency) around the disk center. The sense
of gyration in an elliptic trajectory (clockwise or
counterclockwise) is determined only by the vortex core
polarization. External potentials such as magnetic fields or
(d.c.) spin-polarized currents can also stimulate dramatic effects
such as the switching behavior
\cite{Pulwey01,Volodymyr07,Xiao07,Yamada07}. However, to the
authors knowledge, no internal mechanisms able to induce the core
switching were reported neither experimentally nor theoretically.
Such a chance would only occur if the vortex could interact with
possible inhomogeneities present in the nanostructure. For
instance, a possibility of triggering this process could be
obtained by removing some small portions of the magnetic nanodisk,
in such a way that the cavities (antidots) so created work by
attracting and eventually affecting the vortex structure
\cite{RahmPRL95,RahmAPL82,RahmJAP95,AfranioPRB2005,AfranioJAP2005,LEO,FagnerPLA2004}.
Recently, Rahm and coworkers
\cite{RahmPRL95,RahmAPL82,RahmJAP95,RahmAPL85} have experimentally
studied the cases of one, two, three and four antidots
artificially inserted in a disk with diameter $\sim 500\,{\rm nm}$
and separated by around $150\,{\rm nm}\,-\,200\,{\rm nm}$. Their
results not only confirmed the previous statement about vortex
pinning around the hole defects
\cite{AfranioPRB2005,AfranioJAP2005,LEO,FagnerPLA2004} as well as
have put forward the possibility of using these stable states as
serious candidates for magnetic memory and logical applications
\cite{RahmAPL87}. Although experimental results are provided for
such systems, it is worthy noticing that a broad theoretical
analysis is still lacking. In this paper, we argue that defects
present in the dots may induce interesting possibilities for the
magnetization dynamics, including the switching behavior of vortex
structures. This is a very interesting topic and open new
possibilities for technologies and experiments. Indeed, a very
recent work \cite{Gao08} has shown that magnetization dynamics and
evolution is deeply modified when cavities are introduced. Their
samples consist of Co-made elongated ring-shaped nanomagnets,
which like Permalloy has negligible magnetostatic anisotropy. What
is clearly observed in these experiments is that the reversal
pathway  of the so-called diagonal onion state is drastically
affected by the inserted holes once they attract and sometimes pin
the transient vortex-like domain-wall configurations. Actually,
not only the presence, but also the sizes and locations of the
cavities (two or four in their hollowed samples) remarkably change
magnetization evolution and eventually lead to its reversal. We
will follow a different line in which the magnetization reversal
takes place due to the effects produced by the hole on the vortex
core motion. No further agent is necessary in the switching
process, like occur to other usual methods where external fields
are one of the main ingredients.

Before starting our analysis, it would be useful to describe how a
magnetic vortex can arise in nanodisks. Indeed, a small revision
will help to justify the magnetic model used here. The vortex
state in magnetic nanodot materials is the result of the
competition between exchange and magnetostatic interactions.
Particularly, in finite systems like magnetic dots, densities of
``magnetic charges" are induced in their volumes and surfaces.
Considering the magnetization $\vec{M}$, the volumetric and
superficial densities are defined as
$\rho=-\vec{\nabla}\cdot\vec{M}$ and $\sigma=\vec{M}\cdot \hat{n}$
respectively, where $\hat{n}$ is the unit vectors normal to the
surfaces of the dot at each point. Therefore, in order to minimize
the magnetostatic energy, the magnetization tends to point
parallel to the dot surfaces and it is one of the properties
responsible for the formation of the vortex ground state. However,
as long as we go towards the vortex center, exchange energy
density gradually increases like $r^{-2}$. Then, in order to
regularize exchange contribution, the magnetic moments around the
center tend to revolve, developing out-of-plane projection, so
that exactly at the vortex center it is perpendicular to the disk
face, whose direction defines what is called vortex polarization
($p=+1$, if the centered moment points {\em up}, or $p=-1$, if it
is pointing {\em down}). Here, we study very thin ferromagnetic
nanodisks with thickness $L$ and radius $R$ so that the aspect
ratio $L/R \ll 1$. In this case, one can assume
$\vec{\nabla}\cdot\vec{M}=0$ and hence, the only source of
magnetostatic interactions is the superficial magnetic charges. To
write the vortex state in these dots, it is convenient to
parametrize the magnetization $\vec{M}(\vec{r})$ by two scalar
fields, the polar $\theta(\vec{r})$ and azimuthal $\phi(\vec{r})$
angles, $\vec{\mu}(\vec{r})=\vec{M}(\vec{r})/M=(\sin\theta
\cos\phi, \sin\theta \sin\phi, \cos\theta)$. Using
$m(\vec{r})=\cos\theta(\vec{r})$, the vortex state configuration
can be written as
$m(\vec{r})=m_{0}^{\pm}(\vec{r})=\cos\theta_{0}^{\pm}(\vec{r}),
\phi(\vec{r})= \phi_{0}^{\pm}(\vec{r})=\arctan(y/x) \pm \pi/2,$
where the magnetization unit vector deviates out of the plane at
the vortex core $m_{0}^{\pm}(\vec{0})\rightarrow \pm 1$ and
outside this region, $m_{0}^{\pm}(\vec{r})\rightarrow 0$. The
vortex core size is approximately the exchange length $l_{0}$
(which can be taken to be of the order of the lattice spacing $a$
in our simulations on a discrete lattice). In this notation, the
values of $m_{0}^{\pm}(\vec{0})$ at the vortex center defines the
vortex polarization. On the other hand, $\phi_{0}^{\pm}(\vec{r})$
defines the vortex chirality. The flux closed in-plane vortex
($\phi_{0}^{\pm}(\vec{r})$) and the perpendicular component in the
center ($m_{0}^{\pm}(\vec{0})$) are independent of each other so
that four different magnetization states of one disk are possible.\\

Our paper is organized as follows: in Section II, the model and
associated methods of numerical analysis are presented and
discussed. Section III is devoted to the study of sample with a
single hole. There, we emphasize the vortex core reversal process
induced by its interaction with the defect and the non-ordinary
circumstances favoring it. The case of doubly hollowed disk is
studied in Section IV, where differently from the former case, no
holes configuration yields the core switching: indeed, it is
always captured by one of the defects before any reversal be
observed in our simulations. We finally close our work by point
out our Conclusions and Prospects for
forthcoming investigation.\\
\begin{figure}
\includegraphics[angle=0,width=\columnwidth]{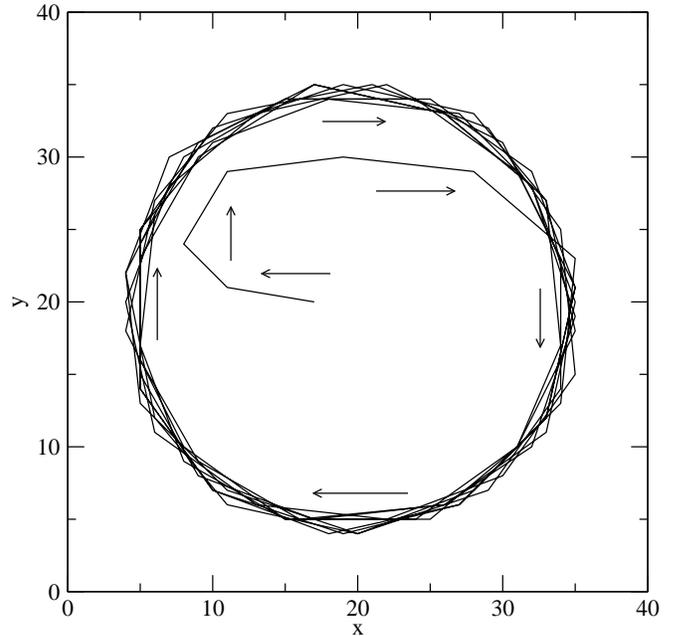}
\caption{ \label{orbita1} Trajectory of the vortex core (beginning
at the disk center $(20a,20a)$), in the $xy$-plane, after turning
off the external field. Here, no hole is inserted and the arrows
indicate the direction of the motion.}
\end{figure}

\section{The model and methods}
In order to describe a thin magnetic nanodisk we pursue an
alternative strategy substituting the disk by a two-dimensional
film with magnetization vectors distributed in a regular square
lattice inside a circumference of radius $R$. In addition, the
magnetostatic interactions due to the presence of the magnetic
charges in the lateral and top surfaces of the dot are replaced by
local potentials. Since we are going to study systems with one and
two holes and also with no hole, the model used is summarized by
the following general Hamiltonian:
\begin{eqnarray}\label{Exchange-magnetostatic-holes}
H &=& -\sum_{\{i,j\}}J_{ij}\vec{\mu}_{i}\cdot \vec{\mu}_{j} +
\sum_{\alpha=1,2}\sum_{k}\lambda_{\alpha}(\vec
\mu_{k}\cdot\hat{n}_{k,\alpha})^{2} \nonumber\\&& +\sum_{D=1,2}
\sum_{l}\lambda_{D,l}(\vec \mu_{l}\cdot\hat{n}_{D,l})^{2}
-\sum_{i}\vec{h}\cdot\vec{\mu}_{i},
\end{eqnarray}\\
where $J_{ij}=0$ for sites inside the holes and $J_{ij}=J>0$ for
the remaining sites of the film. Here,
$\vec{\mu}_{i}=\vec{M}_{i}(\vec{r})/M_{s}=
\mu_{i}^{x}\hat{x}+\mu_{i}^{y}\hat{y}+\mu_{i}^{z}\hat{z}$ is the
unit spin vector at position $i$ ($M_{s}$ is the saturation
magnetization), the sum $\{i,j\}$ is over nearest-neighbor spins,
the terms with positive constants $(\lambda_{\alpha},
\lambda_{D,l})$ mimic the magnetostatic energies (see below) in
the top face of the disk, $\lambda_{1}$, while $\lambda_{2}$ accounts
for the lateral edge. At the holes borders the analogous constants
taking into account the surface charges at the edges are $\lambda_{1,l}$
and $\lambda_{2,l}$, for holes $1$ and $2$ respectively. For the case without
defects, $\lambda_{1,l}=\lambda_{2,l}=0$ while for the situation
with a single defect $\lambda_{2,l}=0$. In turn, $\vec{h}$ is an
external magnetic field. The hole defects are introduced by
removing a number of neighbor spins around sites
$\vec{r}_{1}=(0,d_{1})$ and $\vec{r}_{2}=(0,-d_{2})$ from the
system ($d_{1}>0$ and $d_{2}>0$). Of course, the number of
neighbor spins removed around a particular position ($\vec{r}_{1}$
or $\vec{r}_{2}$) defines the hole sizes $\varrho_{1}$ and
$\varrho_{2}$. Now, the magnetostatic energy favors the spins
to be parallel to the film surfaces. Therefore, the local unit
vectors $\hat{n}$ defined on the points characterizing the
surfaces  of the material are always perpendicular to these
surfaces (or boundaries): $\hat{n}_{k,1}=\hat{z}$ is perpendicular
to the disk face, in the xy-plane; $\hat{n}_{k,2}$ is radially
perpendicular to the circumference envelop, with radius $R$, at
each point; $\hat{n}_{1,l}$ and $\hat{n}_{2,l}$ are radially
perpendicular to the border points of holes $1$ (pointing to
$\vec{r}_{1}$) and $2$ (directed to $\vec{r}_{2}$),
respectively.

\indent As we have already mentioned, in principle, our model,
Hamiltonian (\ref{Exchange-magnetostatic-holes}),
could describe a very thin disk with aspect ratio $L/R \ll 1$. In
fact, several known experimental results with nanodisks can be
qualitatively reproduced using it. For instance, the vortex core
precession mode (which has a rather low eigenfrequency \cite
{Novosad05,Guslienko+06}) can be easily obtained, at least
qualitatively, with this simple model (see below). However, it is
not well known how defects can change this picture. Defects must
be naturally present \cite{Compton06} or artificially inserted
\cite{RahmJAP95,RahmAPL85,Rahm03} in the film samples. It has been
predicted theoretically
\cite{AfranioPRB2005,AfranioJAP2005,LEO,Pereira05} and also
observed experimentally \cite{RahmJAP95,Rahm03} that small holes
incorporated into the magnetic structures can attract and capture
magnetic vortices. On the other hand, we argue that another effect
can occur: the vortex core switching due to a involved dynamical
state of the magnet, which should include the coherent
magnetization oscillations (near the defect) matched to the vortex
motion. Actually, a somewhat similar effect has been observed in
quite recent experiments \cite{Gao08}. There, the
pathway developed to reverse the diagonal onion state in elongated
nanorings is quite sensitive to intentionally inserted cavities,
as well as their sizes and locations. Here, the picture comprises
a vortex-like configuration with polarization, say, up, which
after interacting with a hole, of a given size at a specified
position (these values clearly depend on other parameters like the
whole nanodisk size, applied field amplitude for resonantly
exciting the gyrotropic mode, etc), turns it to be down,
eventually reversing the sense of its gyrotropic motion. What
should be stressed is that such a remarkable phenomenon show up in
our simulations only for very special situations involving the
physical parameters, for instance, the hole must be relatively
small (around 5\%-10\% of the sample size). In addition, the
insertion of a second similar hole in any position deeply
jeopardize the magnetization evolution and the vortex core is
always captured by one of the holes before any core reversal
process.

\begin{figure}
\includegraphics[angle=0.0,width=\columnwidth]{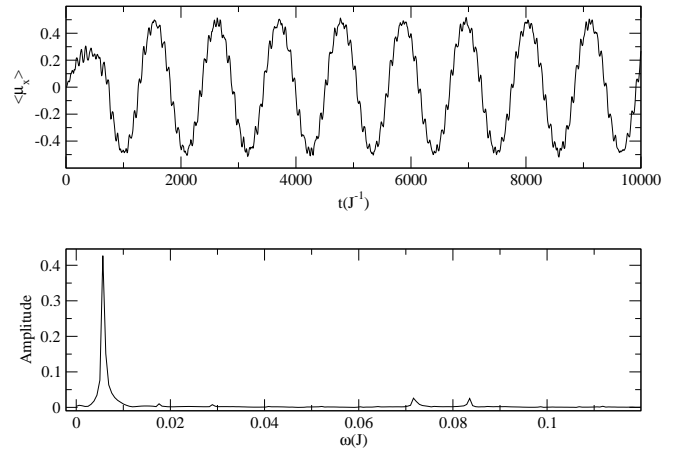}
\caption{ \label{mxsb}  Magnetization in the $x$-direction as a
function of time and its Fourier transform.}
\end{figure}

\begin{figure}
\includegraphics[angle=0.0,width=\columnwidth]{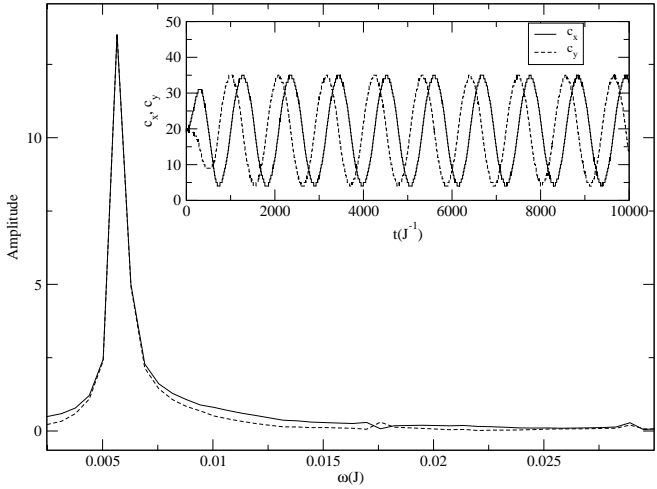}
\caption{ \label{cxcysb} Position components ($c_{x},c_{y}$) of
the vortex center as function of time. Note that the Fourier
transform leads to a peak at the same frequency observed in Fig.\
\ref{mxsb}. This is the resonance frequency of the gyrotropic
motion}
\end{figure}

\begin{figure}
\includegraphics[angle=0.0,width=\columnwidth]{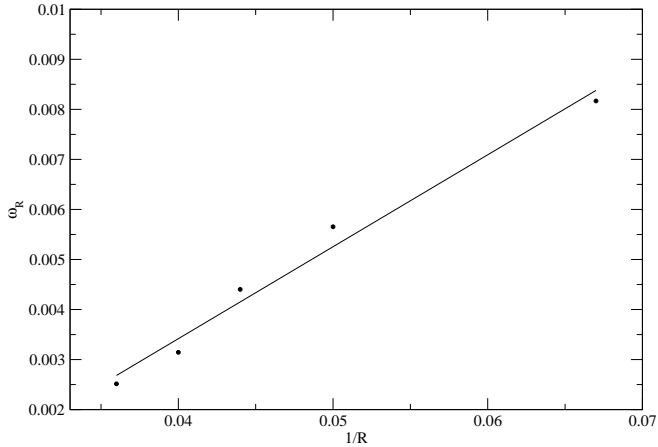}
\caption{ \label{fXR} The behavior of the gyrotropic frequency,
$\omega_R$, as function of the `aspect ratio', $R^{-1}$, obtained
from the present model. Points are simulational results while the
line is the linear fitting.}
\end{figure}

\indent Our results are obtained by using spin dynamics
simulations for a square lattice occupying all possible points
inside a circumference of radius $R$ (in most simulations we have
used $R=20a, 25a, 30a$). It is solved by employing the
fourth-order predictor-corrector method. The calculations
presented here consider only $R=20a$, $\lambda_{1}=0.2J$ and
$\lambda_{2}=2J$. The choice of other values of $\lambda_{1}$ does
not alter the essential physics if it is not large enough (in all
investigations we have used the range $0<\lambda_{1}<0.28)$. For
$\lambda_{1}>0.28$, the vortex becomes essentially planar and does
not develop the out-of-plane components at the core
\cite{Wysin94}. On the other hand, we have verified only a limited
range of $\lambda_{2}$. Indeed, the interval $1.8<\lambda_{2}<2.2$
leads to the same quantitative and qualitative results. In order
to excite the vortex core precession mode, an external
perturbation was applied. Here, we have chosen a sinusoidal
external magnetic field of the form $\vec{h}_{ext}(t)=\vec{h}_{0}
\sin(\omega t)$, with $\vec{h}_{0}=0.01J \hat{x}$ and
$\omega=0.089J$. These are the best conditions we have identified
in our simulations favoring the core reversal. For instance, large
changes in these field parameters will not produce the main
phenomena we intend to describe. In all cases, the duration of the
sinusoidal field was $700J^{-1}$ and the total time of the
simulations was $10^{4}J^{-1}$ with time increment equal to
$\Delta t=0.01J^{-1}$. Firstly, in the absence of the defects and
after turning the field off, we observe the precession mode with a
circular trajectory, as expected (see, Fig.\ \ref{orbita1}). For a
better analysis of this mode, we have plotted the behavior of the
magnetization in the $x$-direction $<\mu_{x}>$ as a function of
time and its Fourier transform (see Fig. \ \ref{mxsb}; the same
behavior can be seen for $<\mu_{y}>$). The Fourier transform
presents a peak at a well defined frequency $\omega_{R}=0.0056J$,
which is the frequency of the vortex core precession (the
resonance frequency, see Fig.\ \ref{cxcysb}). Of course, it
depends on the size $R$ of the disk. More realistically, the
resonance frequency depends on the aspect ratio $L/R$. However,
since our model assumes a very thin disk, we can, in principle,
obtain $\omega_{R}$ as a function of $L/R$ only for $L/R<<1$. We
also studied the behavior of $\omega_{R}$ as a function of $1/R$,
which appears to be essentially linear, at least for $1/R<<1$ (see
Fig. \ref{fXR}), in qualitatively accordance with experimental
findings \cite{Novosad05,Guslienko+06}. All above comparisons (and
others not presented here) show that this model works very well.
It gives us a perspective that Hamiltonian
(\ref{Exchange-magnetostatic-holes}) is also able to predict new
facts.

\begin{figure}
\includegraphics[angle=0.0,width=\columnwidth]{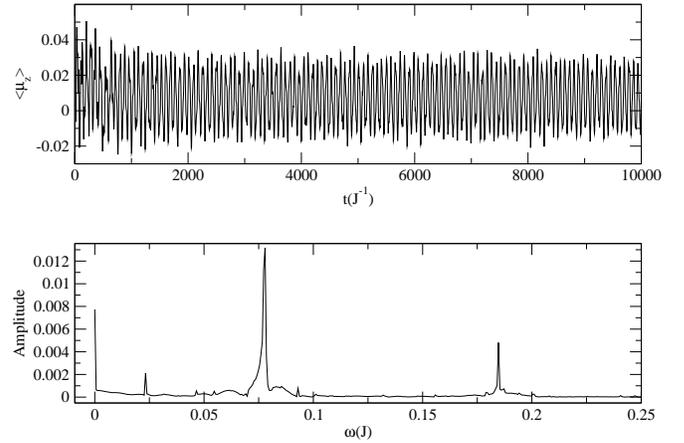}
\caption{ \label{mzsb} $<\mu_{z}>$ versus $t$ and its Fourier
Transform. Note that $<\mu_{z}>$ oscillates around a positive
value.}
\end{figure}

\indent Before studying the problem of interest, we have also
plotted the behavior of the $z$-component of the magnetization as
function of time (and its Fourier transform) in the absence of a
defect (see Fig.\ \ref{mzsb}). In this case, the average
magnetization over the whole sample oscillates around a small
positive value (because the out-of-plane spins forming the vortex
core are pointing along the positive $z$-direction) all the time
without abrupt changes, indicating that there is not any
magnetization reversal during the vortex core precession, as
expected. In addition, the main peak in the Fourier transform
occurs at a frequency $0.078J$, which is different from
$\omega_{R}$.

This model involves several parameters and, of course, it is very
hard to treat in details all aspects. Therefore we have exposed
here the general behavior of the vortex core dynamics and the main
particular phenomena found for specific conditions. We notice that
a hole, independently of its size and position, in general,
attracts the vortex core to its center, i.e., the core almost
always falls into the hole (sometimes, scattering-like events may occur).
Some theoretical
\cite{AfranioPRB2005,AfranioJAP2005,Pereira05,Pereira07,Wysin05}
and experimental works have already reported the capture process
\cite{RahmAPL82,RahmJAP95,Kuepper07}. Hence, it would be
interesting to consider other possibilities for the vortex
dynamics in the presence of such type of defects. Two situations
are analyzed in details. The first one, which is more remarkable,
describes the vortex core switching due to the vortex-defect
interaction. It happens only for very special circumstances and
our work indicates the most suitable configurations, in order to
aid possible experimental probing. The another one relies on the
fate of the core to be trapped by one of the two holes
(two configurations that yield this capture are presented).
Such a situation also serves to illustrate how a relatively small
perturbation eliminates the magnetization reversal and how the
center of the vortex oscillates after a capture process.\\
\section{A nanodisk with a single hole and the vortex core switching}

\indent Now we consider some effects that a hole may induce on the
vortex core dynamics. When a small hole is inserted in the
circular film, the vortex core suffers new influences so that the
magnetization dynamics should become much richer. Indeed, holes
induce an attractive effective potential for the vortex
\cite{AfranioPRB2005,AfranioJAP2005,Pereira05,Pereira07} and
therefore, such force may compete with the effects of applied
fields, magnetostatic, anisotropies, etc.\\
\begin{figure}
\includegraphics[angle=0.0,width=\columnwidth]{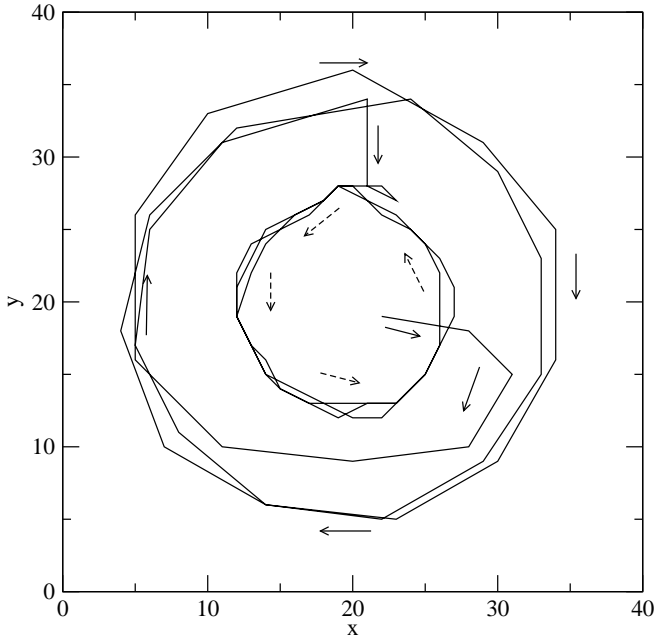}
\caption{ \label{orbita2} Typical trajectory of the vortex core in
a disk containing a hole (at position $(0,33a)$). The vortex core
motion starts in the disk center $(20a,20a)$ and the precession
mode is excited by a field pulse. The comparison of this motion
with that shown in Fig.\ \ref{orbita1} clearly shows the effect
of the hole on vortex core dynamics.}
\end{figure}
\begin{figure}
\includegraphics[angle=0.0,width=\columnwidth]{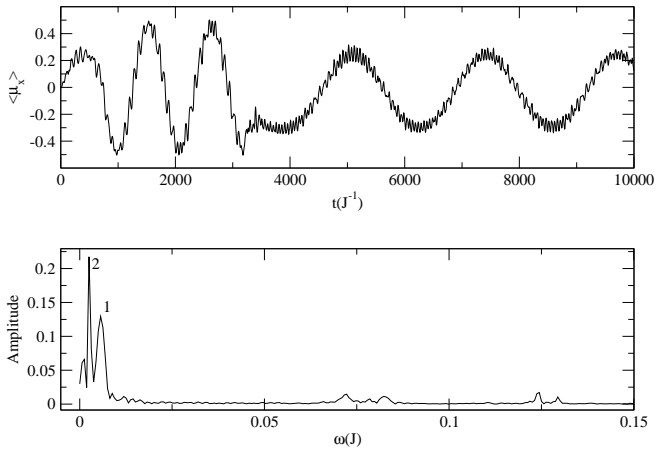}
\caption{ \label{mx} $<\mu_{x}>$ versus $t$ and its Fourier
transform for a film with a hole.}
\end{figure}
\begin{figure}
\includegraphics[angle=0.0,width=\columnwidth]{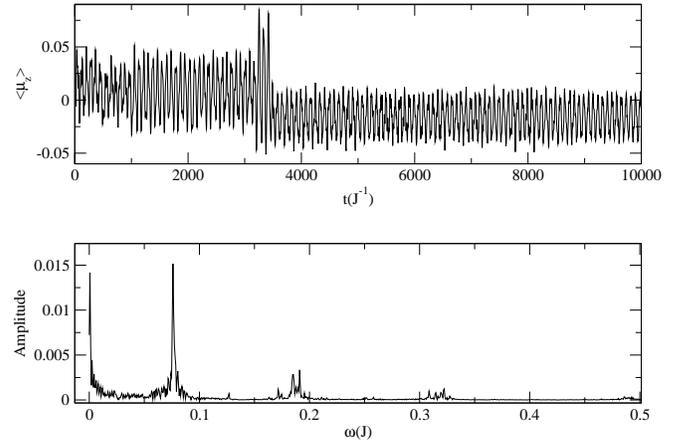}
\caption{ \label{mz} Magnetization in the $z$-direction and its
Fourier transform. The vortex-hole interaction causes the
switching.}
\end{figure}
\indent In this Section we consider the presence of a single hole of
radius $\varrho_{1}=\varrho$. For the external field, we use the
parameters of the previous Section. In a nanodisk with a cavity,
one expects that magnetic charges are also induced in the cavity
walls analogously as they appear in the usual surfaces of the
disk. Therefore, taking into account only the presence of defect
$1$ at position $\vec{r}_{1}$, in principle we should use
$\lambda_{1,l}\neq 0$, which corresponds to the influences of the
magnetostatic energy coming from the edge of the hole cavity. Here
we will show only influences of very small defects as compared
with the disk size and consequently, $\lambda_{1,l}$ could be
neglected in a first approximation since the hole wall can support
only a tiny amount of magnetic charges. More realistic analysis
should then take this parameter to be small, for example, as compared
with $\lambda_{2}$. However, it is illustrative to
study the system for a large range of $\lambda_{1,l}$ values,
say, $0\leq\lambda_{1,l}\leq \lambda_{2}$, for elucidating its
effects on the whole properties of the magnetization. Generally, for the
smallest hole considered (four neighbor spins removed,
$\varrho_{1}=\varrho=a$), the vortex core is almost always
captured (or sometimes scattered) by the defect for
$\lambda_{1,l}> 2.3\times 10^{-4}$. Hence, large enough
contributions of magnetostatic energy coming from the cavity wall
(i.e., large $\lambda_{1,l}$) leads to the common phenomena of
capture, or even the more seldom scattering-like effect, observed for
larger holes. Indeed, larger holes always capture the vortex core,
independently of $\lambda_{1,l}$ (recall that besides magnetostatic,
exchange gets lower as the hole size increases), what agrees with
experimental\cite{RahmAPL82,RahmJAP95,Kuepper07} and theoretical\cite{Pereira07,WAMM08}
results, which consider defects occupying an appreciable fraction
of the disk. Therefore, if one wishes looking for different
possibilities, perhaps small defects should be the focus. With
this strategy, we first analyse several
parameter possibilities and later we shall consider $\varrho=a$ and
use $\lambda_{1,l}= \lambda_{2}/10^{4}$ (or, for simplicity,
$\lambda_{1,l}=0$) for all subsequent calculations. Indeed, the
most important phenomenon we would like to describe here happens
only for $0\leq\lambda_{1,l}\leq 2.3\times 10^{-4}J$. Physically,
it is reasonable to think that a small cavity wall has only a
tiny contribution to the magnetostatic energy.

The size and position of the hole can be controlled and so we have
studied several cases. For example, for a central hole with size
$a$, the resonance frequency of the gyrotropic mode is given by
$0.0034J$, which is smaller than the one we have observed for the
usual nanodisk $\omega_{R}=0.0056J$ (a fact qualitatively
predicted by analytical treatment \cite{WAMM08}). The problem with
this configuration is that the ground state is a vortex pinned to
the hole (of course, without an out-of-plane core) and, therefore,
the vortex needs to be previously released from the cavity to
exhibit the gyrotropic mode. Changes in the defect size and
position lead to the following general situations: depending on
the initial conditions (external sinusoidal fields, essentially),
then, as the vortex core moves sufficiently close the defect
border, it can be either transmitted or captured by the cavity.
These are interesting effects due to core-hole interaction but,
for very special features of the film, a much more rare dynamical
event may be triggered. Actually, choosing the coordinate system
in which the disk center is put at $(20a,20a)$ and if and only if
we use $\vec{r}_{1}=(0,33a)$ (or $(0,7a)$ ) for the center
position (along the $y$-axis at $d_{1}=\pm 13a$) of a hole with
size $\varrho=1a$ (four neighbor spins removed), then a fabulous
phenomenon occurs: the magnetic vortex core is reversed. For all
disk sizes studied, only one configuration was found to be able of
reversing the core magnetization. Always, the hole must be
relatively small and placed near enough the disk border,
perpendicularly to the field. Particularly, for $R=20a$ and
$\rho=a$, the asymmetric conditions with $d_{1}=13a$ seems to be
the unique defect position possible to trigger the switching
process. It should be emphasized that the switching of magnetic
domains generally depends on a myriad of detailed features of the
magnetic particles, and topological effects ultimately limit this
possibility. Therefore, the occurrence of such a delicate
dynamical effect should not be expected for any ordinary
configuration of the magnet with a hole. Usually, what is observed
is the capture of the core when the defect is large enough;
conversely, only small changes in its dynamics are observed for
very small defects. The main lesson is: the core reversal seems to
demand relatively small hole (in our case around $5\%-10\%$ of the
nanodisk dimension) combined with other favorable events, like its
location and suitable applied oscillating field. Another important
point is that, assuming $\lambda_{1,l}/\lambda_{2}<<1$ and for
special conditions and parameters (like $d_{1}$ and $\varrho$ used
here), the reversal of the magnetization of the vortex core (along
the $z$-direction) can be always reproduced. In Fig.\
\ref{orbita2} we show the trajectory of the vortex core. Note that
this mode starts with the same frequency of the case without a
hole (Fig.\ \ref{orbita1}) in a trajectory outer the hole.
However, after some revolutions, when the vortex core moves again
very near the hole border (in a region in between the hole and
disk walls), it quickly changes the direction of the motion
simultaneously with the magnetization reversal along the
$z$-direction (see video of the vortex core motion and the
switching phenomenon as auxiliary material \cite{EPAPS}). Really,
a change in the sense of gyration unambiguously indicates a change
in the vortex core polarization. Then the vortex core again
develops a new precession mode with a smaller frequency
$\omega^{(2)}_{R}=0.0025J$ as shown in Fig.\ \ref{mx} (an orbit
with radius smaller than $d_{y}$). An appreciable part of the
kinetic energy of the vortex was transformed in spin waves. To see
more clearly the switching behavior, the z-component of the
average magnetization $<\mu_{z}>$ is plotted in Fig.\ \ref{mz},
which can be compared with Fig.\ \ref{mzsb}. Note the rapid change
of the mean value of $<\mu_{z}>$ from positive (when the core
direction was up) to negative (when the core is down). In the
Fourier transform, the main peak happens in almost the same
frequency for the case without a hole.
\begin{figure}
\includegraphics[angle=0.0,width=\columnwidth]{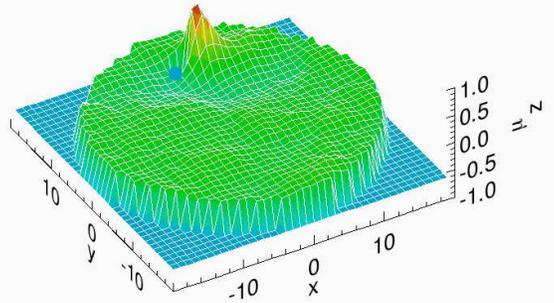}
\caption{ \label{ricardo3d} (Color online) The instant immediately
before the switching process. The out-of-plane magnetization
fluctuations can be perceived in the irregularities present around
the green surface. Note the deep protuberance with negative values
of $\mu_{z}$ near and below the hole and the core. Such a
protuberance will pull the vortex core, leading to the inversion
of its polarity.}
\end{figure}

\indent Thus, the lacking of experimental observation of core
switching induced by vortex-hole interaction may be credited to
the large holes generally inserted into the samples, around
$15\%-25\%$ of its size. In these cases the common phenomenon is
the capture of the core by the defect
\cite{Rahm03,RahmJAP95,Kuepper07}). Basic physics concepts may
help to understand this even better: as it is known the effective
potential (exchange + magnetostatic) experienced by the vortex is
globally minimized inside the hole
\cite{AfranioPRB2005,AfranioJAP2005,Pereira07,WAMM08}. Conversely,
if the hole is very small, it only slightly affects vortex
dynamics and no appreciable effects are expected to occur.
Actually, the phenomenon of core reversal predicted here takes
place for a hole of intermediary size, situation in which the
minima provided by the exchange and magnetostatic contributions
are comparable. In summary, our simulations as a whole agree with
experimental observations upon vortex core dynamics in the
presence of intentionally inserted cavities: generally, the core
is captured by one of the defects, mainly when they are
sufficiently large, like in the experiments of
Refs.\cite{Rahm03,RahmJAP95,Kuepper07}. Only for special
circumstances the core reversal is expected to occur, demanding
among other a non-centered relatively small hole. Another
peculiarity of core reversal induced by vortex-hole interaction is
the fact that the polarization switching occur only from internal
dynamical events: the vortex must be only under motion (the
gyrotropic is the more appropriate for it is easily reproducible)
and pass suitably close to the defect, so that the attraction from
the hole ultimately dominates its inertia, reversing its motion
sense. This in turn, demands the switching in its polarization, so
that the sign of the gyrotropic factor in the
Landau-Lifshitz-Gilbert equation, $\vec{G} \times \vec{v}$ is kept
unaltered. Another dynamical explanation for this phenomenon is:
firstly, the vortex motion excites spin-wave modes. Differently
from the case without defects, the presence of the hole makes
these modes with larger amplitudes due to reflections in its
border, mainly when the vortex core moves near it. Secondly, for a
magnet of finite size, one expects that the radiation of
spin-waves, their reflection off the disk (and hole) boundaries
and their effect back on the vortex core will result in the
establishment of a dynamical state of the magnet which includes
both the confined moving vortex and the coherent magnetization
oscillations matched to the vortex motion. Then, under this
special situation, strong out-of-plane spin fluctuations arise
near the hole border, changing coherently the direction of the
vortex core magnetization when it passes close to the defect.
Figure \ \ref{ricardo3d} shows an instant of the core motion
immediately before the magnetization reversal.

\section{A nanodisk with two holes}
Our aim in this Section is to consider the effect of two holes on
the vortex core dynamics in circular thin films. Also, this study
will be useful for showing that the reversal magnetization
phenomenon is completely suppressed if a second defect, even small
and distant from the first, is introduced into the system. In
addition, the pinned vortex dynamical behavior is detailed. Like
the former case, the disk radius is selected to be $R=20a$.
\begin{figure}
\includegraphics[angle=-90.0,width=7.5cm]{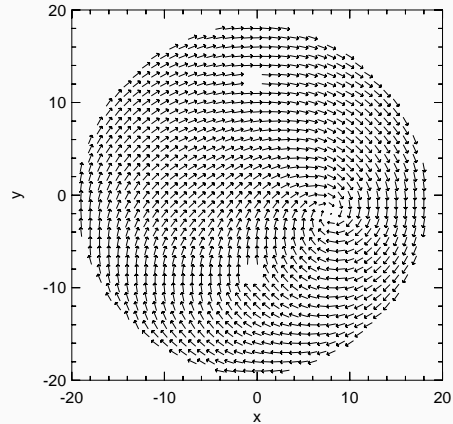}
\caption{ \label{bidimensional1} An instantaneous position of the
vortex core in a circular magnetic thin film with two holes
centered at $(0,13a)$ and ($0,-8a$). The holes sizes are of the
order of the lattice spacing $a$ (in each defect, there are four
spins removed). The length of the arrows are proportional to the
spin projection into the XY-plane. Note the small arrows around
the vortex center at $(8a,-2a)$ forming the vortex core.}
\end{figure}

\begin{figure}
\includegraphics[angle=0.0,width=9cm]{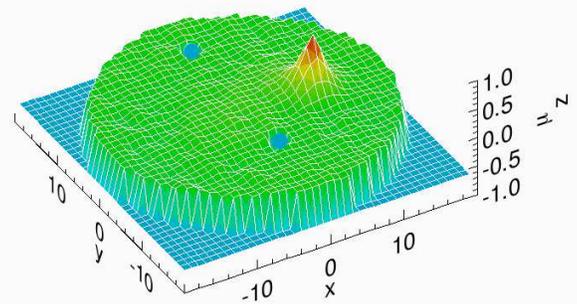}
\caption{ \label{tridimensional2} (Color online) Three-dimensional
view of Fig. (\ref{bidimensional1}). The vortex polarization is up
(yellow-red peak). Note the out-of-plane fluctuations generated by
the core motion. These spin waves propagate through the system and
are reflected in the disk and holes borders.}
\end{figure}
\begin{figure}
\includegraphics[angle=0.0,width=7cm]{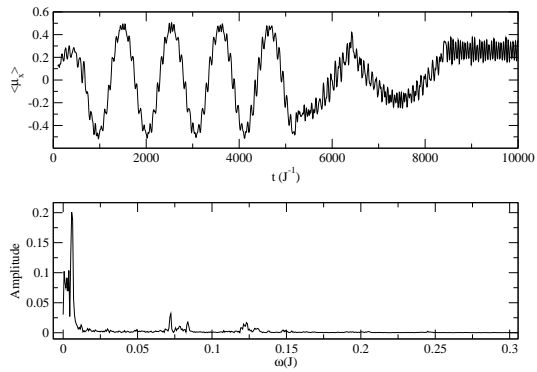}
\caption{ \label{magx} Mean magnetization $<\mu^{x}>$ as a
function of time and its Fourier transform. The resonance
frequency is approximately $0.0057J$. This figure shows clearly
that the vortex core is captured by one hole at $t\sim
8000J^{-1}$.}
\end{figure}
\begin{figure}
\includegraphics[angle=0.0,width=7cm]{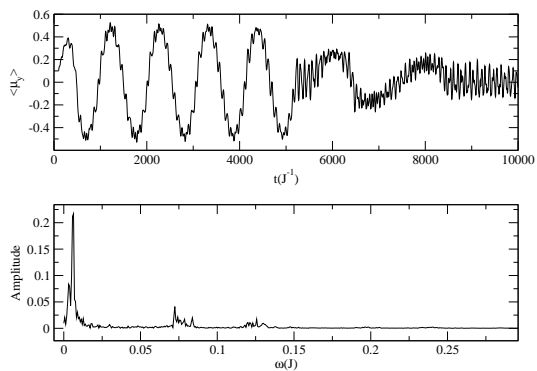}
\caption{ \label{magy} Mean magnetization $<\mu^{y}>$ as a
function of time and its Fourier transform.}
\end{figure}
\begin{figure}
\includegraphics[angle=0.0,width=7.5cm,height=7.5cm]{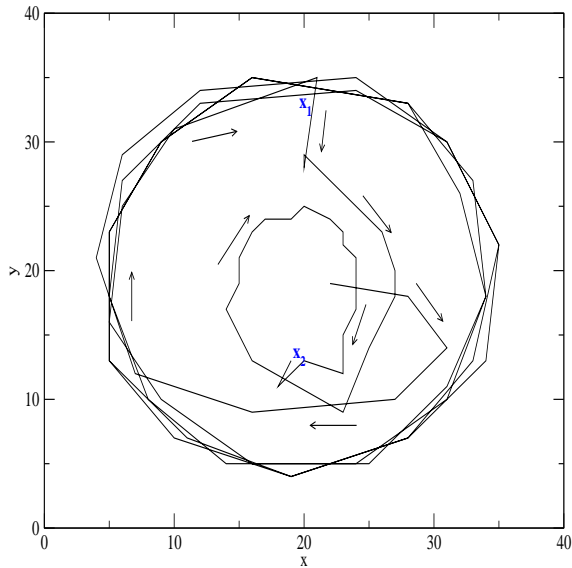}
\caption{ \label{orbitaRod} The complete orbit of the vortex core
until capture by hole $2$. The vortex starts its motion in the
disk center at $(20a,20a)$. The symbols $X_{1}$ and $X_{2}$
indicate the positions of antidots $1$ and $2$ respectively. In
the studied process, the core follows an almost perfect gyrotropic
mode, completing five laps around the film before feeling a strong
effect of hole $1$, which changes its motion direction.}
\end{figure}

The results are obtained by choosing an external sinusoidal field
to excite the gyrotropic mode. Again, the field is applied for a
short time ($\sim 700J^{-1}$) and its parameters are
$\vec{h}_{0}=0.01J \hat{x}$ and $\omega=0.0093J$. Of course, the
vortex motion will depend on the holes size and position as well
as the external field. In all observed cases, the vortex finishes
trapped in one of the antidots just it finds the defect in its
way, even for smaller holes. Since this is the common phenomenon
for all possible positions of the defects, we intend to know the
dynamical process after the trapping. To illustrate this, we
choose the following geometry: hole $1$ at position
$\vec{r}_{1}=(0,13a)$ and hole $2$ at $\vec{r}_{2}=(0,-8a)$ as
shown in Figs. (\ref{bidimensional1}) and (\ref{tridimensional2}).
Larger defects will capture the core even faster. What should be
stressed is that the presence of the second defect, even if
smaller and distant of the first, will change the vortex dynamics
drastically so that no core switching have been observed before
its capture by one of the holes. Figure (\ref{magx}) shows the
mean magnetization $<\mu^{x}>$ along the $x$-direction as a
function of time. As the vortex core moves, this mean
magnetization oscillates almost harmonically from $-0.5$ to $0.5$,
indicating that the motion is approximately circular (or
elliptical) around the disc center. The orbit is outer the holes.
However, at about $t\sim 5000J^{-1}$, the core decreases
considerably its velocity near the hole $1$, moves slowly in
spiral (outer the antidot $2$ and inner antidot $1$), until
hitting antidot $2$ again, where it is captured at about
$t=8000A^{-1}$. Now, the sense of gyration is not changed and
consequently, there is no switching process. The $<\mu^{x}>$
magnetization oscillations stop abruptly becoming almost constant
around $0.4$ (actually, $<\mu^{x}>$ oscillates very rapidly with a
very small amplitude around $0.4$). We also have calculated the
Fourier transform of $<\mu^{x}>$ (see Fig.(\ref{magx})). It has
two main peaks: the first one refers to the oscillating external
field (along the $x$-direction) and the second is related to the
resonance frequency. In our unit system, such resonance frequency
is given by $\omega_{res-d}\sim 0.0057J$, which can be compared
with the analogous problem of a film without defects
$\omega_{R}\sim 0.0056J$. Similar graphics are obtained for
$<\mu^{y}>$ as shown in Fig.(\ref{magy}). The difference is that,
after being pinned, the mean $y$-magnetization oscillates around
zero (with a relatively large amplitude as compared to its
counterpart in $x$) and not around a finite value as happens to
$<\mu^{x}>$. In addition, there is no the first large peak
exhibited in the $<\mu^{x}>$. The reasons are simple: first, the
holes were displayed in the $y$-axis (at $(0,13a)$ and $(0,-8a)$)
and second, the external sinusoidal field (responsible for the
first peak in the Fourier transform of $<\mu^{x}>$) is applied
along the $x$-direction and hence, it can not appear in
$<\mu^{y}>$. The trajectory of the vortex core is shown in
Fig.(\ref{orbitaRod}).

Based on the above description, an interesting fact to report is
that the vortex center oscillates around the hole center with
small amplitude and large frequency when the vortex is pinned to
the defect. This phenomenon looks like the one theoretically
predicted in layered ferromagnetic materials \cite{Pereira05}.
Nevertheless, this effect must be very richer in nanodisks because
the surface magnetostatic energy has an effective contribution,
which is absent in the layered systems. Analytical estimate
\cite{WAMM08} predicts that such oscillations are related to the
vortex mass and take place around $10^5$ GHz, which is far beyond
current possibilities of observation in these systems
\cite{Kupper}, $\sim 10$GHz. However, a study considering all
microscopic details of the pinned vortex oscillation will require
further investigation.\\

\indent In our present analysis, the oscillation frequency of this
mode can be seen in the Fourier transform of $<\mu^{x}>$ or
$<\mu^{y}>$ and is given by $\varpi\sim 0.123A$. It is much higher
than the resonance frequency, and much lower than mass-like
vibrations, as above. Such an intermediary frequency could be
somewhat related to the {\em fractional gyrotropic} mode, which is
predicted to occurs around $10^2 \omega_R$ and is provided by a
remanent (fractional) gyrotropic vector of the vortex core,
partially captured by the hole\cite{WAMM08}. Another interesting
feature can be easily seen when one compares the Fourier
transforms of $<\mu^{x}>$ and $<\mu^{y}>$. Note that, after being
captured, the magnetization amplitude oscillations are larger in
the $y$-direction than in the $x$-direction. It means that the
pinned vortex center oscillates much more along the $x$-axis than
along the $y$-axis. Therefore, it describes an ellipse with larger
semi-axis along the $x$-direction, of the order of the hole size.
For a trapped vortex, the difference in the amplitudes of
oscillation of $<\mu^{y}>$ and $<\mu^{x}>$ increases as the hole
responsible for the capture is placed nearer the circumference
border of the film. Hence, this ellipse turns out to be an almost
straight line if the hole is located very near the disk boundary,
while it takes a circular shape if the cavity is centered in the
disk. Therefre, several procedures of production and propagation
of spin waves with different phases could be built only by
changing the defect position, which is easily manipulated by
lithographic techniques. Thus, our results indicate that the
vortex-hole system could also works as a mechanism to create and
control spin waves in confined magnetic thin films.
\begin{figure}
\includegraphics[angle=0.0,width=7.5cm]{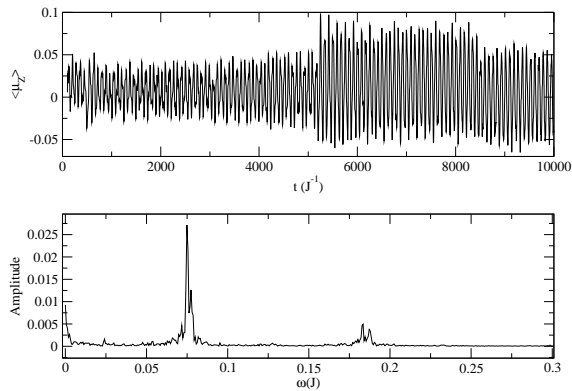}
\caption{ \label{magz} Mean magnetization $<\mu^{z}>$ as a
function of time and its Fourier transform.}
\end{figure}

We have also plotted the mean magnetization along the
$z$-direction, $<\mu^{z}>$, and its Fourier transform,
Fig.(\ref{magz}). There, we clearly realize that, before the
capture process, $<\mu^{z}>$ oscillates around a small positive
value, once the vortex is {\em up}-polarized (negative if the
polarity were {\em down}). After some vortex core laps, at about
$t\sim 5000J^{-1}$, the amplitude of oscillation increases
considerably. This is just the moment the vortex core motion
suffers a strong disturb (near hole $1$) and starts to slow down,
changing the orbit until $t\sim 8000J^{-1}$, when the core is
captured by hole $2$. Therefore, the kinetic energy of the core
decreases considerably and a large amount of spin waves is
released during the instant of the capture process. The pinned
vortex oscillations also generate more and more spin waves due to
the rapid rotations of vortex center. The amplitude of $<\mu^{z}>$
remains large after trapping but now its oscillations are quite
small, meaning that the pinned vortex polarization practically
vanishes.
\begin{figure}
\includegraphics[angle=0.0,width=8cm]{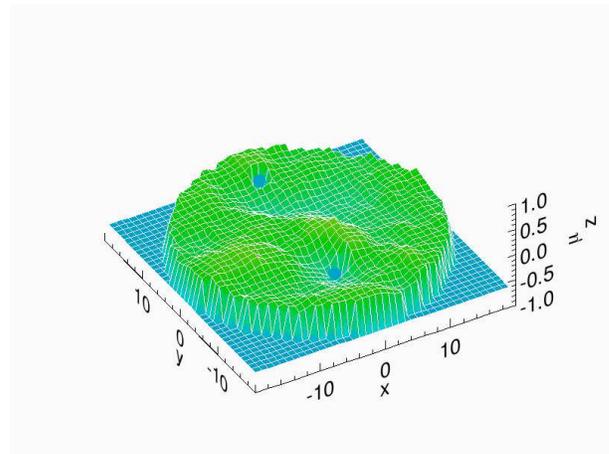}
\caption{ \label{3dburaco} (Color online) A picture of the instant
that the vortex core was captured by hole $2$. Note the large
amount of spin waves generated in this process. Such waves can be
seen by the out-of-plane fluctuations given by irregularity in the
spin pattern (green surface) on the film.}
\end{figure}\\

\begin{figure}
\includegraphics[angle=0.0,width=\columnwidth]{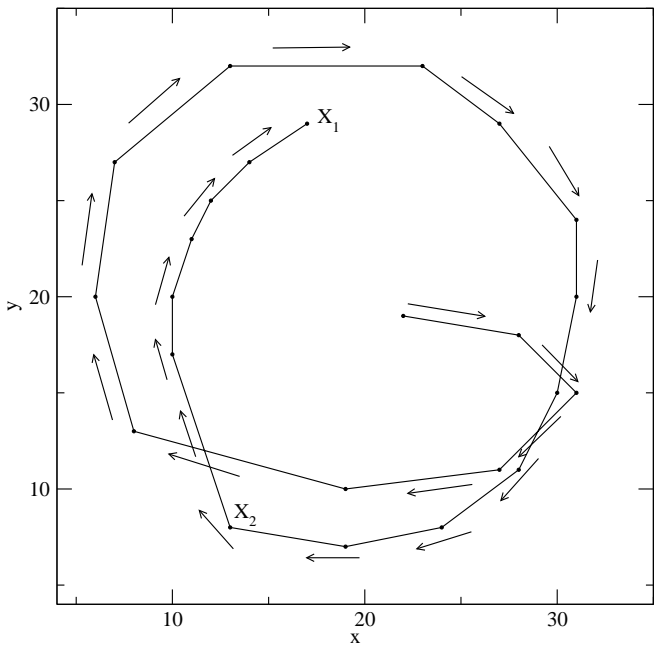}
\caption{ \label{orbita2holes} The orbit of the vortex core in a
nanodisk with radius $R=20a$ centered at $(20a,20a)$ containing
two holes at positions $X_{1}=(0,33a)$ and $X_{2}=(12a,8a)$. As
observed for several geometries with two defects, the fate of the
core is to be captured by one of the holes.}
\end{figure}

 \indent For the sake of completeness, we now discuss a situation
 where the holes are not aligned along the same line crossing the
 disk center. Let us consider a disk with radius $20a$ centered at
 position $(20a,20a)$ with two small holes ($\varrho_{1}= \varrho_{2} =
\varrho = a$) displayed at $(0,33a)$ and $(12a,8a)$.
Figure (\ref{orbita2holes}) shows the orbit of the vortex core
until its capture by one of the holes. Even in this situation
(many others, with the holes closer and apart from each other
were also tested) the main facts remain
the same: after being excited by the applied sinusoidal field the
vortex core starts its gyrotropic motion around the disk center.
Whenever passing near a hole its dynamics changes (similarly to
the former case studied, say, both holes centered along $y$-axis,
as depicted in Figures \ref{tridimensional2}, \ref{magx}, and
\ref{magy}), and after some revolutions it is eventually captured
by one of the defects. Analogously to the previous configuration
with two holes, no core reversal was observed in our simulations,
neither for smaller (only $1$ spin removed) or larger ($8$ spins
removed) defects. However, we cannot strictly state that no other
sizes could trigger core reversal (with two or more holes), once
their sizes in our simulations can be only largely varied (keeping
its circular shape), say, while the smallest hole is obtained by
removing a unique spin, the next possible size demands we take 4
spin away, and so forth. The main point must be recalled: if
core reversal could somewhat occur with two or more holes, the
simplest system where this phenomenon would certainly be easiest
and clearest observed is that comprising a nanodisk with a single
non-centered relatively small hole.

\section{Conclusions and prospects}
In this paper we have considered an alternative model for circular
magnetic thin films containing artificial defects and we have also
predicted an interesting way of reversing the magnetization of the
vortex core which is completely dependent on its dynamics and
interaction with a relatively small hole, inserted in the sample.
Therefore, it should be stressed that the present switching
process is not stimulated by external agents, like usually occur
by means of the application of magnetic or spin-polarized DC
current fields. The conditions necessary are summarized by the
following properties: first, only one hole must be present and
localized at a position near the lateral border of the disk; we
suggest that the center of the hole must be placed at a distance
on the order of the vortex core size away from the disk border.
Second, the hole must be small and third, the field required to
excite the gyrotropic mode should be applied perpendicularly to
the line joining the disk and hole centers.[Although the field is
an external agent, we emphasize once more that its role is
restrict for resonantly excite the gyrotropic mode, putting the
vortex core in motion. Such a field has no direct influence on the
reversal mechanism, which is completely triggered by vortex-hole
interaction]. Clearly, the suitable
parameters of the applied magnetic field, relative hole size, etc,
should be probed in actual experiments.\\

\indent As we observed in our simulations, those special
conditions force the core to move in a circular trajectory passing
exactly in between the hole and disk borders generating a large
amount of spin waves, which will also develop important role in
the switching process (for example, see Fig.(\ref{ricardo3d}).
Nowadays, such defects were already intentionally introduced in
nanostructures \cite{RahmJAP95,RahmAPL85} but they occupy a
relatively large fraction of the material (about $15$-$25$
percent). Hence, the switching mechanism occurring during the
vortex motion can be met only by further miniaturization of
defects in nanostructures. Also in this line, we should mention
the very recent experiment\cite{Gao08} where hole defects induce
magnetization reversal in elongated $Co$ rings. The results are
aimed at the same direction of our work and are an experimental
indication that our proposal may have further and broad relevance.
An important difference between our prediction and the phenomenon
experimentally observed in Ref.\cite{Gao08} is that the last needs
a permanent presence of an external magnetic field (while in our
case the mechanism of the vortex switching is based only on
internal interactions during the vortex motion). In this aspect,
the dynamical characteristic of the phenomenon predicted here is
another differential with relation to the usual core reversals
induced by external potentials.

To achieve more details about the vortex-hole interactions, we
have also considered a system with two holes. In this case no
switching process was observed for several configurations
analyzed. Since one of the holes will trap the vortex, we have
then addressed our attention to the magnetization dynamics after
the capture. When the vortex center falls into a hole, some of its
energy is radiated away in the form of spin waves. The vortex
center oscillates inside the hole with a frequency much larger
than the resonance frequency. The bound state vortex-hole releases
a large quantity of spin waves forming coherent magnetization
oscillations matched to the vortex structure (part of the kinetic
energy of the core lost in the capture is transformed in spin
waves). These results demonstrate a significant coupling between
the vortex (with the
core or even captured, without the core) and spin waves in a disk.\\

To conclude we would like to suggest that the switching mechanism
reported here can be used in technological applications. Actually,
this phenomenon naturally lends itself to applications in binary
data storage. Therefore, it seems that our investigations may
provide fundamentally new ways of using magnetic nanostructures in
technology.\\

\indent The authors thank CNPq, FAPEMIG and CAPES (Brazilian agencies) for
financial support.

\end{document}